# Adaptation of TURN protocol to SIP protocol

**Mustapha GUEZOURI, Ahmed BLAHA and Mokhtar KECHE**

**Department of Electronics, Faculty of Electrical Engineering, University of Science and Technology (USTO), P.O. Box 1505 – El-m'naouar, Oran, ALGERIA**

**Abstract**

Today, SIP is a protocol par Excellence in the field of communication over Internet. But, the fact that it belongs to the application layer constitutes a weakness vis-à-vis the NAT traversal. This weakness is due to the way in which the server replies to the requests of clients on the one hand. On the other, it is caused by the dynamic allocation of UDP ports for emission and reception of packets RTP/RTCP. The TURN Protocol may face this weakness. However, its use requires a certain number of exchanges between the clients and a TURN server before establishing the multimedia sessions and this increase the latent time. In this article, we propose to adapt TURN protocol for applications based on SIP protocol such as telephony over Internet, conference video, etc. This adaptation optimises the establishment of multimedia sessions by integrating a manager of TCP connections and multimedia flow controller into SIP Proxy server.

## 1. Introduction

The convergence of communication systems of companies over IP networks (Internet Protocol) [1] seems an inevitable phenomenon. A multitude of protocols and standards have been developed to facilitate this convergence such as: H.323 [2], MGCP (Media Gateway Control Protocol) [3], and SIP (Session Initiation Protocol) [4]. Among all the proposed standards, SIP protocol dominates the field thanks to its manageable use and facility of integration in the platforms.

SIP is certainly the current leader but not a miracle protocol. However, the fact that it belongs to the application layer causes two major problems with the network address translator (NAT) [5]. The first one relates to signaling while the second has to do with multimedia sessions, or more precisely their description by the clients located behind the NAT. One of the proposed solutions is the TURN protocol (Traversal Using Relay NAT) [6].

## 2. Network Address Translator

Presently, the NAT (*Network Address Translation*) could not be circumvented in the majority of network topologies; whenever a network is connected to other networks, its use is required. Its utility is not only restricted to economizing the necessary IP addresses to connect for example a private network to Internet. It is also concerned with security: it offers the protection of the machines from external attacks.

Four types of NAT are defined in [7] and [8]:

### 2.1 Full cone

It is the simplest of NAT; it does not impose any restriction. All the packets coming from outside may cross this NAT if they are received on a port which has been already used for the emission of a packet coming from a local machine. Example: a machine with IP address: 192.168.1.11 which tends to receive and send packets through port 5600. The NAT translates this private IP address into a public IP address as well as the number of port, 192.168.1.11:5600 → 68.92.25.44:4325; any user on Internet tending to send a packet to 68.92.25.44: 4325, this packet will be transmitted to the machine with IP address: 192.168.1.11 on port 5600 by the NAT.

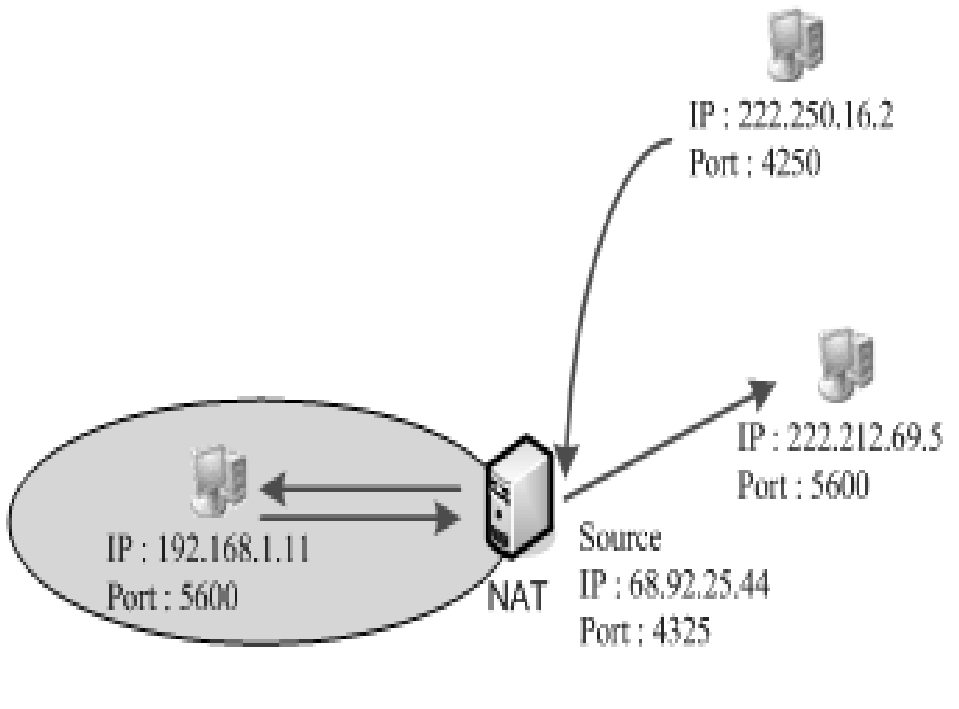

Fig. 1 Full cone NAT

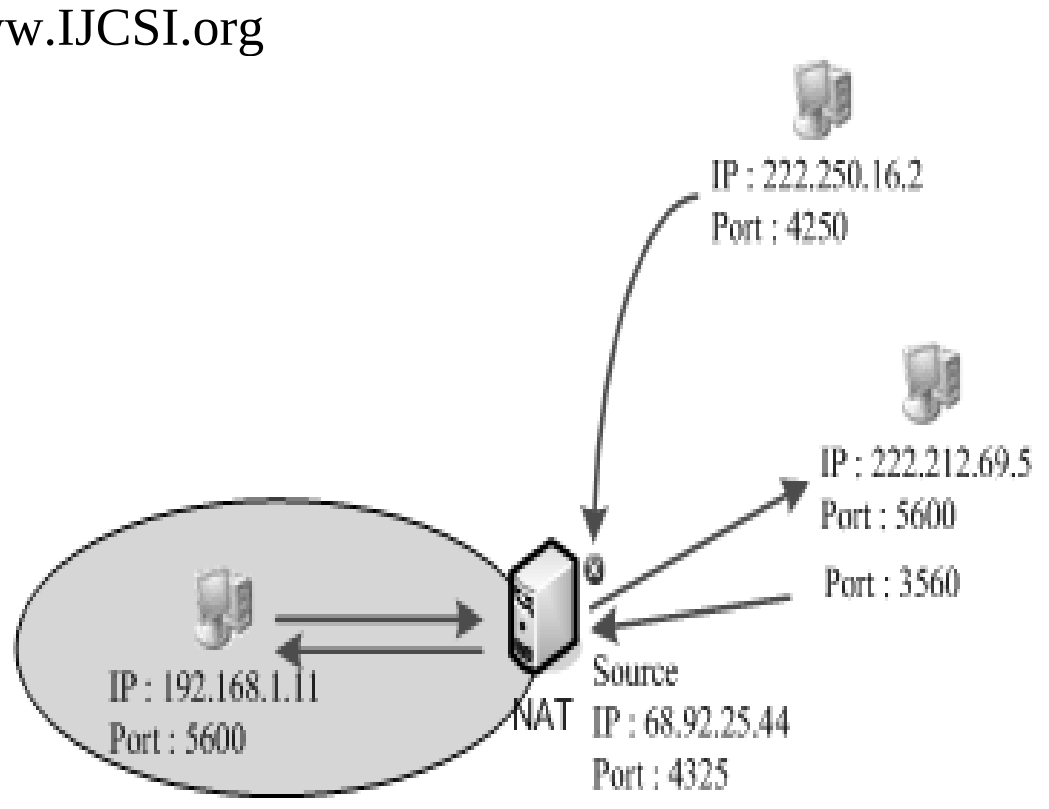

Fig. 2 Restricted cone

## 2.2 Restricted cone

In this case, only the packets received on a port from where one or more packets have been sent to the IP source of the received packets will not be blocked, i.e. The peripheral NAT maintains a mapping between the (private IP address, private port) and (NAT's IP address, NAT's port). If the latter example is taken (Figure 2), the NAT will block the packets sent from the address 222.250.16.2. On the other hand if a packet is sent from the address 222.212.69.5 and even if the port is different from 5600 it will be accepted and transmitted to the local machine.

## 2.3 Port restricted cone

It is identical to the restricted cone, but in this case the NAT blocks all the packets coming from an IP address and the different port to which the local machine has already sent packets.

Example, see Figure 4, if a packet is sent from 222.212.69.5: 3560 it will be blocked because the local machine has sent a packet to port 5600 and not to port 3560.

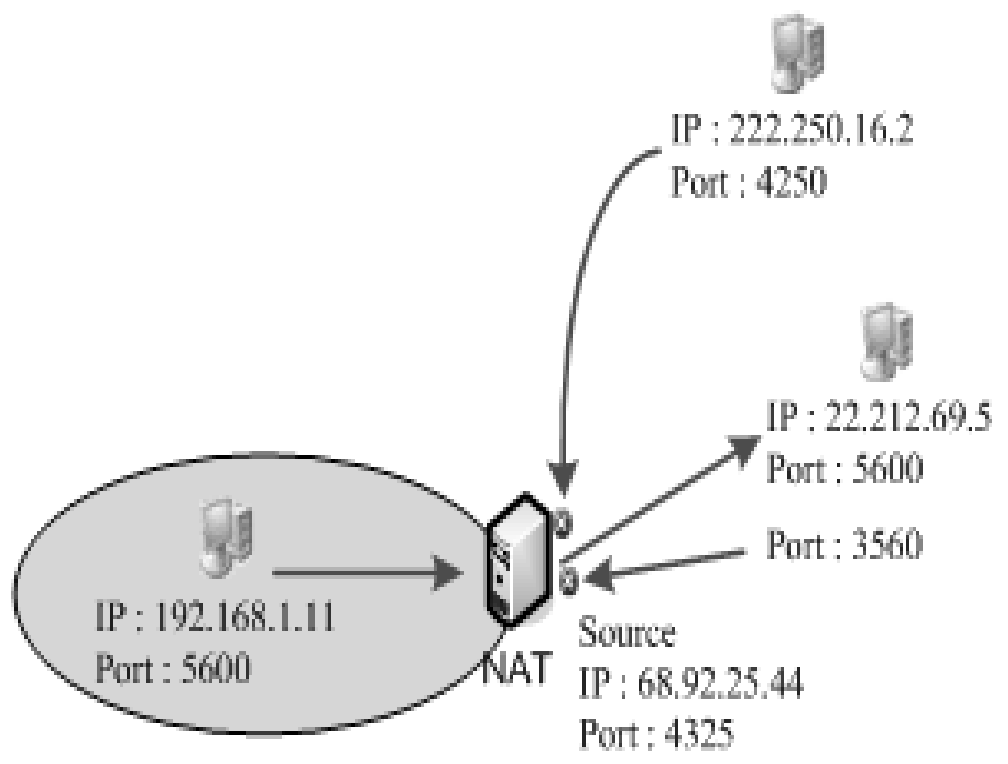

Fig.3 Port restricted cone

## 2.4 Symmetric cone

It is the most complex type of NAT. Each time the local machine sends a packet to a new pair (IP address: port number) a new translation is made, i.e. a machine could communicate by using the same private port source (on Figure 4 port 5600) with several other external machines by using various pairs (IP address: port number).

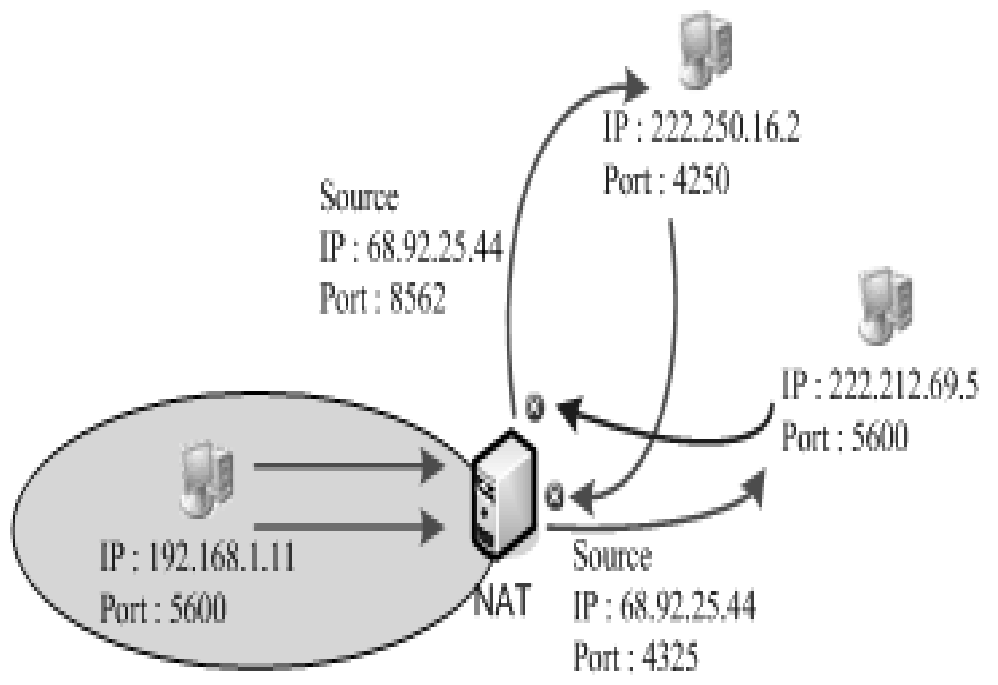

Fig. 4 Symmetric cone

## 3. Session Initiation Protocol (*SIP*)

A call of telephony on Internet is composed of two parts: a signalling part based on a protocol of signalisation like SIP and a media stream part in which the protocol RTP (Real-time Transport Protocol) is used [9]. This is to encapsulate the packets containing the voice and RTCP protocol (Real-time Transport Control Protocol) [9] for control.

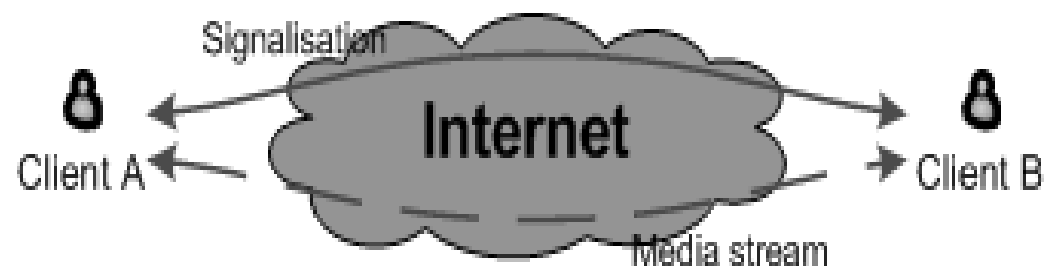

Fig.5 Telephony over Internet

## 3.1 SIP signalling

To transmit the requests and responses, the SIP protocol also borrows UDP (User Datagram Protocol) [10] such as TCP (Transport Control Protocol) [11]. This has been since its second version. Let's take the case when UDP is used as the protocol of transport and show where the difficulty with NATs lies in. Figure 6 illustrates the registration phase of a client behind a NAT with the Registrar Server.

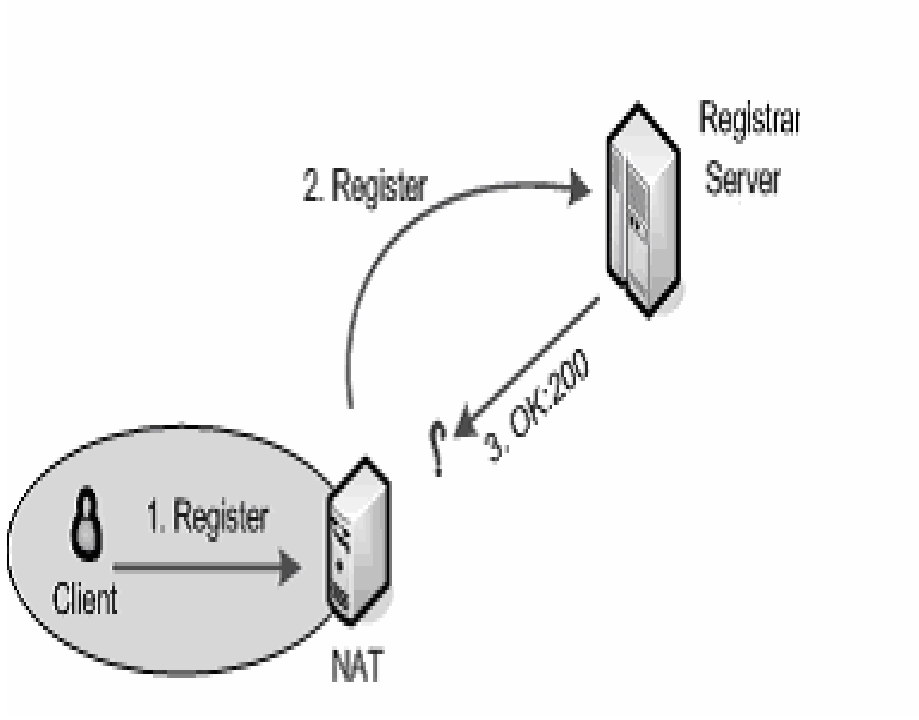

Fig. 6 Registration phase

The problem is due to the way the server replies to the client's request. In fact, it is based on the IP address and port number contained in the VIA field. However, the value of this field contains a private IP address of the client who is behind a NAT. Consequently, the response of the server will not arrive to the right destination and the problem is the same for all the other types of requests. A solution has been already proposed, it consist in telling the server about the client who is behind a NAT. This is by the addition of a parameter in the field VIA by the client himself [12]. The server, identifying this parameter, will reply according to the IP address located in the heading IP and the number of port located in the heading UDP. The later relates to the datagram which contains the request instead of IP address and the number of port located in the VIA field. However, this solution is insufficient. Indeed, if the server receives an Invite request destined for client A after a time T, such that T > 60 s, this request will be blocked at the level of NAT. This is because generally the mapping concerning UDP ports is refreshed every 60 seconds at most. i.e. a NAT closes a UDP port which has been opened if it does not receive anything on this port at the end of a time T, such that 0 < T < 60 S.

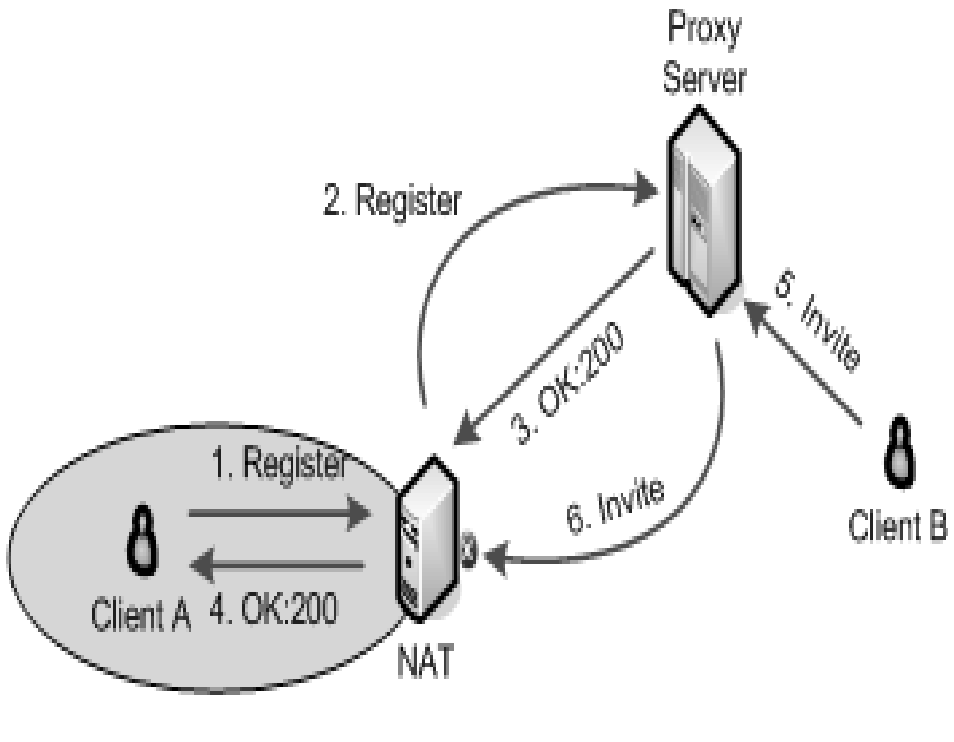

Fig.7 Signalisation and NAT

As a solution, a periodic emission of the Register requests by clients could be proposed to the server in order to refresh the mapping. But, this periodic emission tends to increase the traffic of network and may saturate it.

Some people may think that TCP is enough to solve the problem. Although, it is not the case. SIP, in fact, does not require the server to maintain the various TCP connections open to the clients. Therefore, it is often based on the IP address and number of port contained in the field VIA. Thus the problem remains.

## 3.2 Media stream

In media stream, RTP protocol is used to encapsulate the voice and UDP protocol is borrowed for transport. The problem here relates to the dynamic allocation of UDP ports in order to send and receive the packets containing the voice. Figure 9 is a good illustration of this problem.

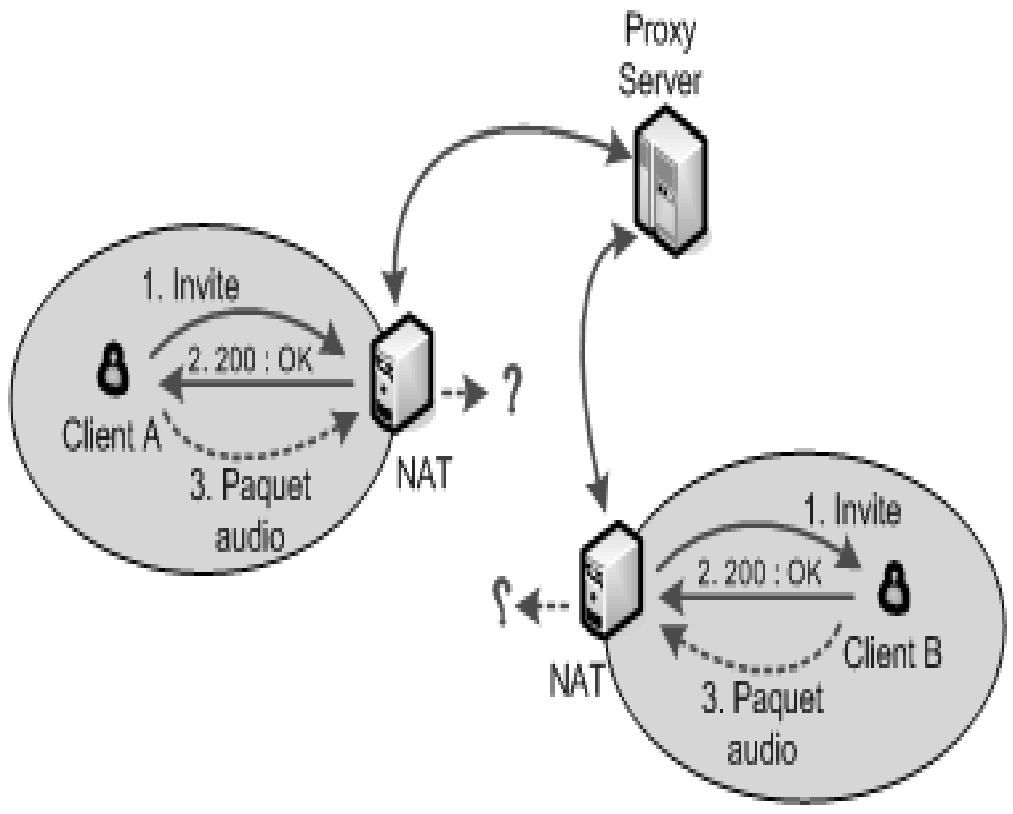

Fig. 8 NAT and media stream

According to the example of Figure 8, two clients behind two distinct NATs tend to establish a call. Let us suppose that the signalling part occurred correctly. As soon as the two clients start sending their packets of voice, the later will not be correctly transported. This is due to the information described in the body of Invite request (Figure 9) and in that of response OK:200 (Figure 10) using SDP (Session Description Protocol ) [13]. This is as far as the ports of reception of each client are concerned. They are valid only in their local area network. So, solving this problem requires the two clients to know the mapping which will be used if they tend to receive the packets of voice on the ports described in the request Invite and response OK: 200.

```
INVITE sip:ClientB@local2.com SIP/2.0

Via: SIP/2.0/TCP 192.168.1.11
From: ClientA <sip:ClientA@local1.com>
To: ClientB <sip:ClientB@local2.com>
Call-ID: 12345625400@local1.com
CSeq: 1 INVITE
Contact: ClientA <sip:Client@192.168.1.11>
Content-Type: application/sdp
Content-Length: 147

v=0
o=ClientA 2890844526 28902245844526 IN IP4 local1.com
s=Session SDP

c=IN IP4 192.168.1.11

t=0 0

m=audio 49570 RTP/AVP 0

a=rtpmap:0 PCMU/8000
```

Fig. 9 The Invite request

```
SIP/2.0 200 OK
Via: SIP/2.0/TCP 192.168.1.11;branch=z9hG4bK77ef4c2312983.1
 ;received=45656465446464
From: ClientA <sip:ClientA@local1.com>
To: ClientB <sip:ClientB@local2.com>
Call-ID: 12345625400@local1.com
CSeq: 1 INVITE
Contact: <sip:ClientB@10.0.0.4>
Content-Type: application/sdp
Content-Length: 131

v=0
o=ClientB 284586526 28922265 IN IP4 local2.com
s=Session SDP

c=IN IP4 10.0.0.4

t=0 0

m=audio 6580 RTP/AVP 0

a=rtpmap:0 PCMU/8000
```

Fig. 10 The 200: OK response

Among the solutions suggested to solve the problem of NATs: STUN protocol (Simple Traversal of UDP through NAT) [14], connections oriented-media [15] and TURN protocol. TURN is distinguished from the other solutions by its capacity to give a transparent traversal vis-à-vis NATs in all the possible situations and with any type of NAT. Traversal Using Relay NAT (TURN) is a protocol that allows to elements behind a NAT or firewall to receive incoming data over TCP or UDP connections. It is even used in cases where a symmetric NAT takes place [6]. The idea is to give to each element an IP public address and port number, which makes it accessible by outside. The allocation of

One IP address and a port number is fulfilled through the exchange of a certain number of TURN requests and responses. Therefore, if the case of telephony over Internet is considered with SIP as a signaling protocol, 2 requests for allocation will prevail: one for the reception of SIP messages and the other for the reception of RTP/RTCP packets [9]. This generates heavy establishment of calls.

To overcome face this slowness, we propose an adaptation of TURN protocol for the applications using SIP protocol such as telephony over Internet, video conference, etc. This adaptation consists in eliminating the allocation requests and integrating in the *SIP Proxy* server a manager of TCP connections and multimedia flow controller.

## 4. The adaptation of TURN for SIP

Instead of the fact that each client requests the allocation of resources (IP addresses and port number) to a TURN server in order that SIP messages could be received during the signaling phase; we propose the use of TCP protocol as a transport protocol and integration into *SIP Proxy* sever a manager of TCP connections. TCP is used instead of UDP to benefit from a greater period of refreshing of mappings at the level of NAT. The manager of TCP connections plays the role of maintaining all TCP connections initialised by the clients during the registration phase. It also transports correctly to them SIP messages through these connections. Figure 1 illustrates this principle. When the register request of client A gets to *SIP Proxy* server, the manager of TCP connections deals with the TCP connection through which the register request has been emitted. The same is and so is for the register request of client B. Then, client B initialises the call by sending an Invite request to *SIP Proxy* server. On the level of the later it is the manager of TCP connections, which will transmit the request to client A through the TCP connection, initialised during the registration phase. The same procedure remains valid for all the other types of SIP messages exchanged between the two clients.

For the reception of RTP/RTCP packets, the allocation request of resource will be eliminated. This is by forcing the clients to use the same UDP port for reception and emission of their RTP/RTCP packets and by integrating into *SIP Proxy* server a multimedia flow controller. The principal functions of the later will be management of a UDP ports pool, modification of SDP [13] bodies of SIP messages transmitted by clients and correct routing of RTP/RTCP packets to clients. Figure 12 shows the operation of multimedia flow controller, and Figure 13 show the call establishment.

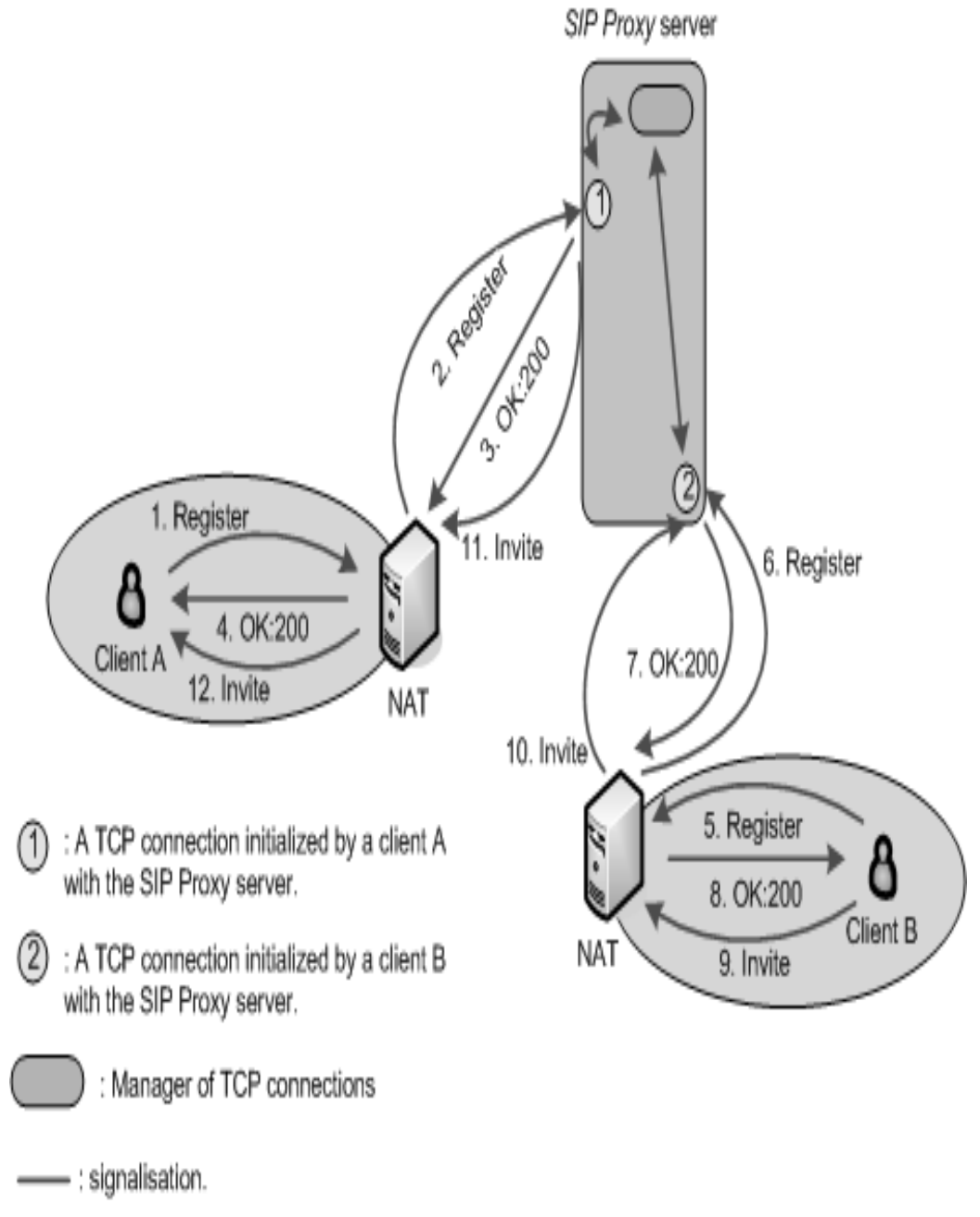

Fig.11 The adaptation of TURN in signaling phase

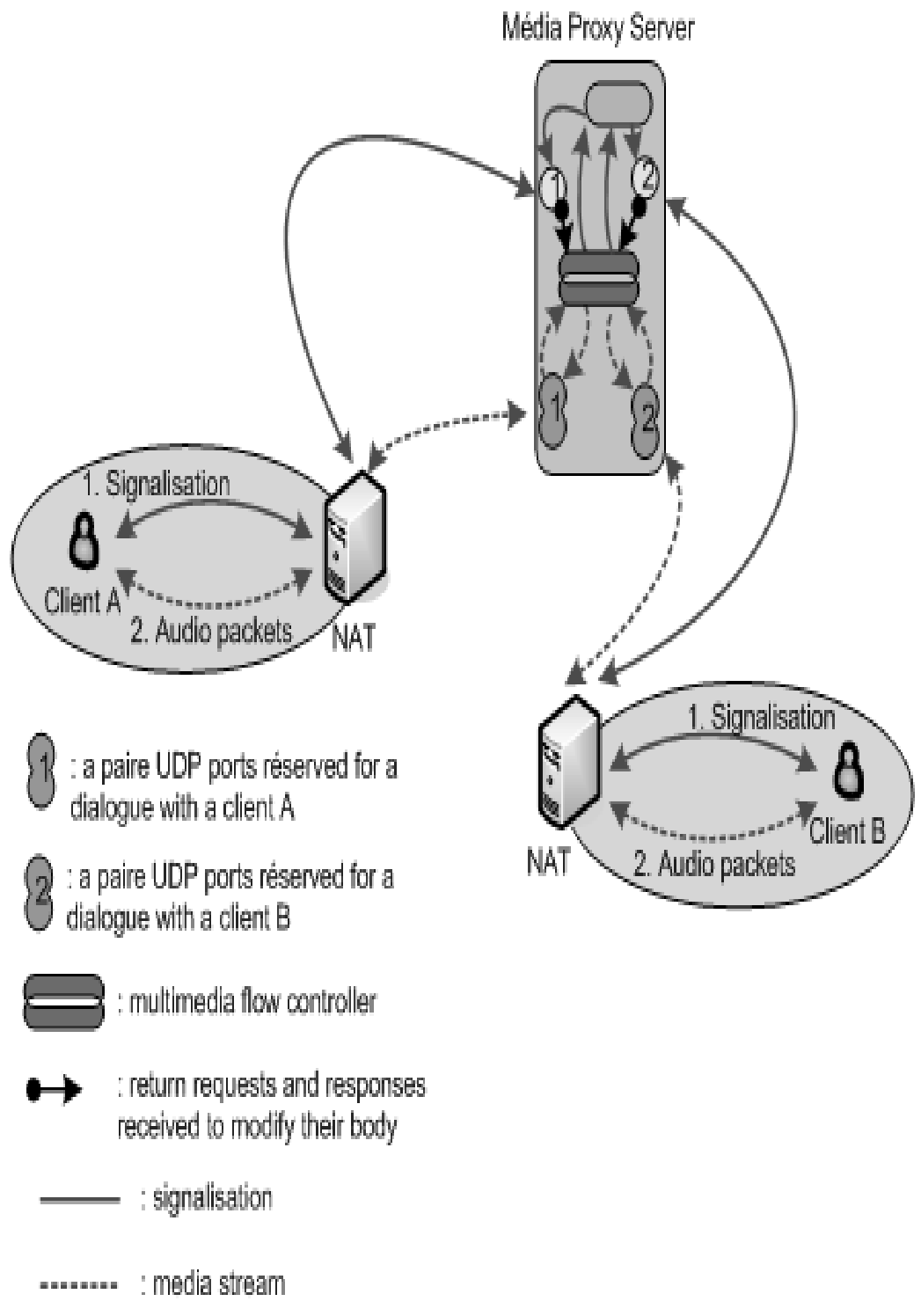

Fig. 12 The adaptation of TURN in media stream phase

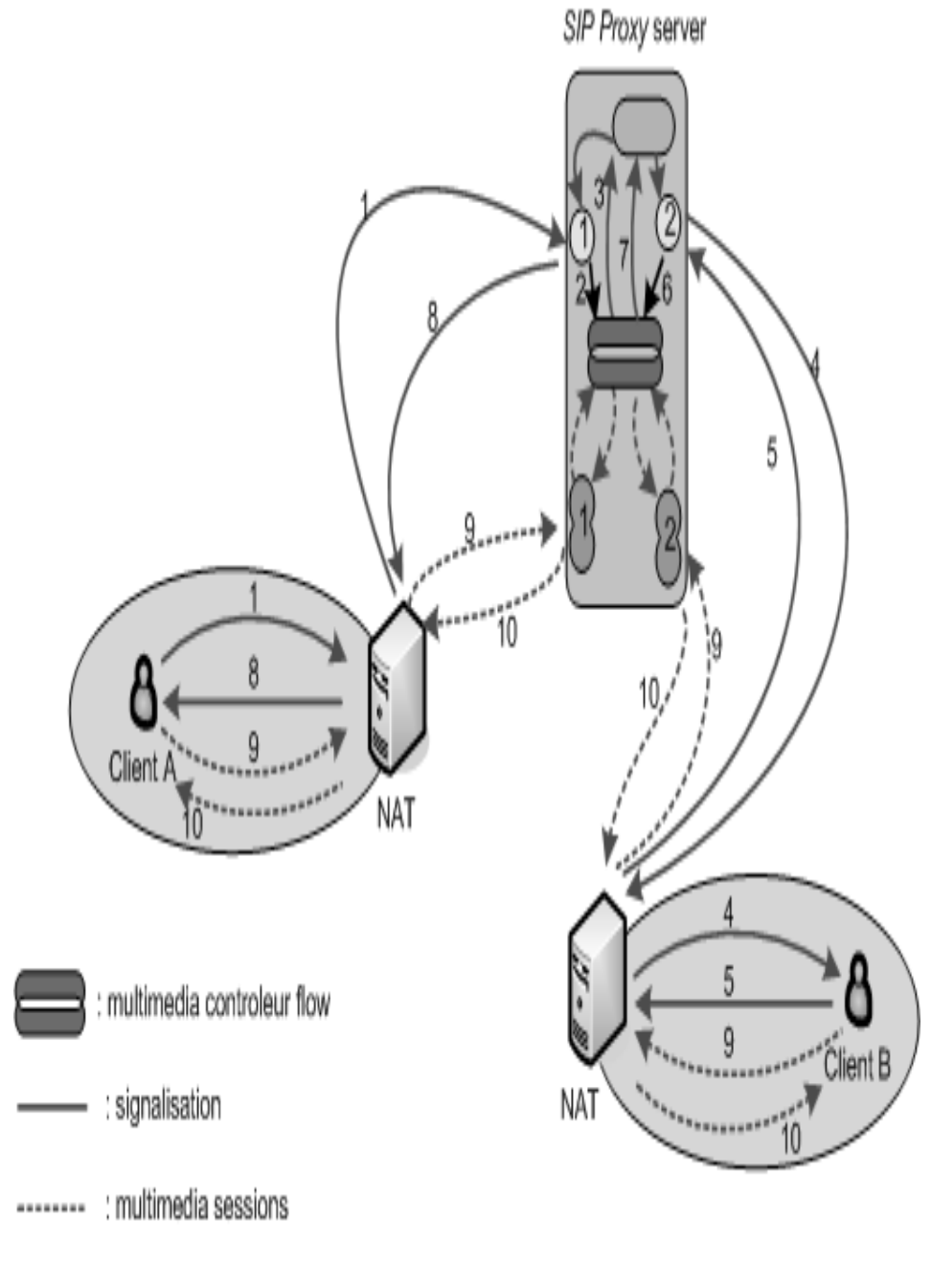

Fig.13 Call establishment

- Client A sends an *Invite* request to *SIP Proxy* Server.
- The *Invite* request is transmitted to the multimedia flow controller, which modifies its body after that UDP ports have been reserved from its port pool in order to receive the RTP/RTCP packets from the client B.
- The modified *Invite* request is transmitted to the manager of TCP connections, which is responsible for transmitting it through the TCP connection, initialised by the client B.
- The new *Invite* request is sent to client B.
- Client B replies with a *200:OK* response that includes the IP address and port numbers which will be used to receive RTP/RTCP packets.
- Client B is also behind a NAT; thus, the IP address and port numbers contained in SDP body of its response are erroneous. The multimedia flow controller should modify this IP address by that of *SIP Proxy* server and recovers from his pool of port UDP necessary ports to replace those mentioned by the client B.
- The modified 200: OK response is transmitted to the manager of TCP connections which is in charge of transmitting it through the TCP connection, previously initialised by client A.

Now, each client knows the number port and address IP to which it should send its RTP/RTCP packets. But, the multimedia flow controller has not determined yet the ports used by the clients to send RTP/RTCP packets because each client is behind a NAT. Therefore; the multimedia flow controller should wait for the arrival of first RTP/RTCP packets.

- Client A sends RTP/RTCP packets to IP address and port numbers received in the body of *200: OK* response. This IP address and these port numbers are those of the *SIP Proxy* Server and so does client B.
- The multimedia flow controller transmits RTP/RTCP packets sent by client A to client B and the same for RTP/RTCP packets sent by client B, they are transmitted to client A.

## 5. Conclusion

The adaptation of TURN protocol proposed in this article gives for SIP protocol a transparent traversal of NAT by optimising the principle of TURN protocol. The later is the allocation of resources, by eliminating the exchanges related to allocation requests. The experiments undertaken on telephony over Internet platform of the PC-to-PC type within the framework of this study allowed showing that the adaptation suggested simplifies the session's multimedia establishment by using the principle of resource allocation but with a completely transparent way.

M.GUEZOURI was born in Algeria in 1962. He received the engineer degree from the national Algerian institute of Telecommunications (ITOran) in 1988.
He received the Master and PhD degrees in electrical engineering in 1990 and 2007, respectively from the Science and Technology University in Oran. Since 1990, he has been a Research Assistant at Signal and processing Laboratory, University in Oran.
He is currently an Assistant Professor of electrical and computer engineering at the University in Oran, Algeria.
His current research interests include Signal processing, neural networks, IP Networks and Ad hoc networks.